\documentclass[twocolumn,amsmath,amssymb]{revtex4}
\usepackage{graphicx}
\usepackage{dcolumn}
\usepackage{bm}
\begin{document}

\title{Spin Transitions in Graphene Butterflies at an Integer Filling Factor}
\author{Areg Ghazaryan and Tapash Chakraborty$^\ddag$}
\affiliation{Department of Physics and Astronomy,
University of Manitoba, Winnipeg, Canada R3T 2N2}

\date{\today}
\begin{abstract}
Recent experiments on the role of electron-electron interactions in fractal
Dirac systems have revealed a host of interesting effects, in particular, the
unique nature of the magnetic field dependence of butterfly gaps in graphene. 
The novel gap structure observed in the integer quantum Hall effect is quite 
intriguing [Nat. Phys. {\bf 10}, 525 (2014)], where one observes a suppression of the
ferromagnetic state at one value of the commensurable flux but a reentrant ferromagnetic 
state at another. Our present work that includes the interplay between the electron-electron 
interaction and the periodic potential, explains the underlying physical processes that 
can lead to such a unique behavior of the butterfly gaps in that system where spin flip 
transitions are involved in the ground state. 
\end{abstract}
\maketitle

The fascinating dynamics of Dirac fermions in graphene has been exhaustively studied in recent
years \cite{graphene_book,abergeletal,FQHE_chapter}. Coulomb interaction between Dirac fermions
\cite{interaction}, in particular, in the presence of a strong perpendicular magnetic field
has resulted in the fractional quantum Hall states in monolayer \cite{mono_FQHE} and bilayer
graphene \cite{bi_FQHE}, which have also been experimentally observed \cite{FQHE_expt}. Graphene
placed on boron nitride with a twist displays fractal butterflies \cite{langbein,hofstadter}
of Dirac fermions \cite{graphene_butterfly}, when subjected to a perpendicular magnetic field. 
After the exciting discovery of the fractal butterflies in graphene \cite{dean_13,hunt_13,geim_13} 
more recent theoretical \cite{apalkov_14} and experimental \cite{geim_14} studies have focused on 
the influence of electron-electron interactions on the butterfly gaps. Given the intricacies of
these gaps, the interaction effects are more complex in the integer and fractional quantum Hall effect 
regime, where one observes an interplay between the quantum Hall effect gap and the Hofstadter gap 
\cite{areg_butterfly}. In studying the interaction effects in the integer quantum Hall effect regime, 
Yu et al. \cite{geim_14} employed capacitance spectroscopy to explore the `Hofstadter minigaps' for 
zero and integer filling factors. Their results for the energy gaps at filling factors $\nu=0, \pm 
1$ $(\nu=n\phi^{}_0/B$, $n$ is the particle density and $\phi^{}_0$ is the flux quantum) showed very 
unusual magnetic field dependence. In the low magnetic field region, the $\nu=0$ gap rises {\it 
linearly} with $B$ and saturates near the magnetic flux value $\phi=\phi^{}_0/2$, but exhibits a 
minimum at $\phi=\phi^{}_0$. In these two regimes, the gap deviates significantly from the Coulomb 
energy thereby indicating that the transitions across the gap from the ground state do not necessarily 
involve the particle charge alone. 

By employing the magnetic translation group algebra \cite{mag_translation} in the quantum Hall
effect regime \cite{haldane_85,read,FQHE_book,areg_butterfly}, we have analyzed the magnetic
field dependence of the $\nu=0$ butterfly gaps. Our results reveal that the observed gap structure
involve spin flip transitions in the ground state, as explained below. We consider graphene in an 
external periodic potential \cite{apalkov_14,review,vidar,ando}
\begin{equation}
V(x,y)=V^{}_0[\mathrm{cos}(q^{}_xx)+\mathrm{cos}(q^{}_yy)],
\label{PotentialForm}
\end{equation}
where $V^{}_0$ is the amplitude of the periodic potential and $q^{}_x=q^{}_y=q^{}_0=2\pi/a^{}_0$, 
where $a^{}_0$ is the period of the external potential. Then the many-body Hamiltonian is
\begin{equation}
{\cal H}=\sum_i^{N^{}_e}\left[{\cal H}^i_B + V(x^{}_i,y^{}_i)\right] +
\tfrac12\sum_{i\neq j}^{N^{}_e}V^{}_{ij}
\label{MBHamiltonian}
\end{equation}
where ${\cal H}^i_B$ is the Hamiltonian of an electron in graphene in a perpendicular magnetic 
field and the last term is the Coulomb interaction. The electron energy spectrum of graphene 
has twofold valley and twofold spin degeneracy in the absence of an external magnetic field, 
the periodic potential and the interaction between the electrons. It is well known 
\cite{graphene_book,abergeletal} that for magnetic fields that are presently accessible in the 
experiments, the conservation of the SU(2) valley symmetry in the presence of the Coulomb interaction 
is a fully justified approximation. We therefore employ this approximation in our studies. In order 
for the external periodic potential to break the SU(2) valley symmetry, the scattering process 
will require momentum transfer comparable to the value of the difference of momentum between the
two valleys. The period of the external potential accessible in the experiment for the
moir\'e superlattice is much bigger than the graphene lattice constant. Therefore the probability 
for such a momentum transfer processes is exponentially small and can be disregarded. In 
order to investigate what kind of state of total valley and real spin is favored by the 
system in our calculations, we consider both the spin and valley degrees of freedom of the 
electron system, where the spin degeneracy is lifted due to the Zeeman effect, while in the 
approximation described above there is no term in the Hamiltonian (\ref{MBHamiltonian}) to 
lift the valley degeneracy. The single-particle Hamiltonian ${\cal H}^{}_B$ is then written 
as \cite{graphene_book,abergeletal,FQHE_chapter} 
\begin{equation}
{\cal H}^{}_B=\xi v^{}_F\left(\begin{array}{cc} 0 & \pi^{}_- \\ \pi^{}_+ & 0 
\end{array}\right)+\frac12 g^{}_\mathrm{e} \mu^{}_B B \sigma^{}_z,
\label{SBHamiltonian}
\end{equation}
where $\pi^{}_\pm=\pi^{}_x\pm i\pi^{}_y$, ${\bm \pi}=\mathbf p +e\mathbf A/c$, $\mathbf p$ 
is the two-dimensional electron momentum, $\mathbf A=(0,Bx,0)$ is the vector potential, 
$v^{}_F\approx 10^6\,\mathrm{m/s}$ is the Fermi velocity in graphene and the last term is 
the electron Zeeman energy. The $\xi$ is the valley index: $\xi=1$ for valley $K$ and 
$\xi=-1$ for valley $K'$. The honeycomb lattice of graphene consists of two sublattices 
A and B and the two component wave functions corresponding to the Hamiltonian 
(\ref{SBHamiltonian}) can be expressed as $(\psi^{}_A,\psi^{}_B)^T$ for valley $K$ and 
$(\psi^{}_B,\psi^{}_A)^T$ for valley $K'$, where $\psi^{}_A$ and $\psi^{}_B$ are wave 
functions of sublattices A and B, respectively. The eigenfunction of the Hamiltonian 
(\ref{SBHamiltonian}) for both $K$ and $K'$ valley can be written in the form 
\cite{graphene_book,abergeletal,FQHE_chapter}
\begin{equation}
\label{BaseEig}
\Psi^{}_{n,j}=C^{}_n\left(\begin{array}{c} \mathrm{sgn}(n)(-i)\varphi^{}_{|n|-1,j} \\
\varphi^{}_{|n|,j}\end{array}\right),
\end{equation}
where $C^{}_n=1$ for $n=0$ and $C^{}_n=1/\sqrt{2}$ for $n\neq0$, $\mathrm{sgn}(n)=1$ for 
$n>0$, $\mathrm{sgn}(n)=0$ for $n=0$, and $\mathrm{sgn}(n)=-1$ for $n<0$. Here 
$\varphi^{}_{n,j}$ is the electron wave function in the $n$-th LL with the parabolic 
dispersion, taking into account the periodic boundary conditions (PBC) \cite{FQHE_book,note}. 
The eigenvalues of Hamiltonian (\ref{SBHamiltonian}) corresponding to the eigenvectors 
(\ref{BaseEig}) for both valleys $K$ and $K'$ are $\epsilon^{}_n=\mathrm{sgn}(n)\hbar\omega^{}_B
\sqrt{|n|}$, where $\omega^{}_B=\sqrt{2}v^{}_F/\ell^{}_0$, $\ell^{}_0=\sqrt{c\hbar/eB}$ 
is the magnetic length.  

We consider a system of finite number $N^{}_e$ of electrons in a toroidal geometry, i.e., 
the size of the system is $L^{}_x=M^{}_xa^{}_0$ and $L^{}_y=M^{}_ya^{}_0$ ($M^{}_x$ and
$M^{}_y$ are integers) and apply PBC in order to eliminate the boundary effects. Defining the 
parameter $\alpha=\phi^{}_0/\phi$, where $\phi=Ba_0^2$ is the magnetic flux through the unit 
cell of the periodic potential and $\phi^{}_0=hc/e$ the flux quantum, we have
\begin{equation}
\frac{N^{}_s}{M^{}_xM^{}_y}=\frac1\alpha,
\label{FluxCond}
\end{equation}
where $N^{}_s$ describes the LL degeneracy for each value of the spin and valley index. In 
this work we consider the filling factor of $\nu=0$, which means that the number of electrons
in the zeroth LL is $N^{}_e=2N^{}_s$, because of the fourfold degeneracy of each LL in graphene. 
The procedure of constructing the Hamiltonian matrix in the basis of the many-body states 
$|j^{}_1,j^{}_2,\ldots,j^{}_{N^{}_e}\rangle$ (besides $j^{}_i$, each single-particle state is 
characterized by the LL, spin and valley indices which are not shown, but are implicitly assumed to
be included in the indices $j^{}_i$) constructed from the single-particle eigenvectors (\ref{BaseEig}). 
Diagonalization of this matrix follows the procedure outlined in \cite{areg_butterfly}.
It was previously shown \cite{areg_butterfly} that the center of mass (CM) translations $T^\mathrm 
{CM} (\mathbf a^{}_p)$ with the translation vector $\mathbf a^{}_p=m\beta^{}_1a^{}_0\hat{\mathbf 
x} + n\beta^{}_2a^{}_0\hat{\mathbf y}$ can be used to characterize the eigenstates of Hamiltonian 
(\ref{MBHamiltonian}), where the eigenvalues of the CM translation operator have the form 
$\mathrm{e}^{2\pi i\left(\beta^{}_1ms/M^{}_x+\beta^{}_2nt/M^{}_y\right)}$. Here, 
$m$ and $n$ are integers which define the translation vector, $\beta^{}_1$ and 
$\beta^{}_2$ are integers which characterize the degeneracy of the many-body system and can be 
determined from the relation $N^{}_e\beta^{}_1\beta^{}_2/\alpha=\pm1,\pm2,\ldots$, $s$ and 
$t$ are also integers which are defined modulo $M^{}_x/\beta^{}_1$ and $M^{}_y/\beta^{}_2$ 
respectively and describe the eigenvalues of the CM translation operator. The CM translations 
allow us to divide the basis states into equivalence classes and transfer the complete 
Hamiltonian into block diagonal form where each block is diagonalized separately and also to
characterize each eigenvalue of the Hamiltonian with the appropriate CM momentum. Therefore 
using the CM translation analysis the Hamiltonian matrix can be approximately divided into 
$M^{}_xM^{}_y/\beta^{}_1\beta^{}_2$ separate blocks. 

In what follows we consider the two cases $\alpha=1$ and $\alpha=2$. We then choose the system 
size based on the condition (\ref{FluxCond}) and the number of electrons. After comparing 
the results for small system sizes for the cases with and without the inclusion of the 
contribution of higher LLs, in what follows we disregard the LL mixing and present all the 
results for the $n=0$ LL. Here we present the results for three system sizes, $N^{}_e=8$ 
taking into account both $K$ and $K'$ valleys, $N^{}_e=6$ and $N^{}_e=8$ taking into account 
only the $K$ valley. For $N^{}_e=8$ in two valleys the system size is $M^{}_x=2$ and $M^{}_y=2$ 
for $\alpha=1$, and $M^{}_x=4$ and $M^{}_y=2$ for $\alpha=2$. For $N^{}_e=6$ in one valley 
the system size is $M^{}_x=3$ and $M^{}_y=2$ for $\alpha=1$, and $M^{}_x=6$ and $M^{}_y=2$ 
for $\alpha=2$, and for $N^{}_e=8$ in one valley the system size is $M^{}_x=4$ and $M^{}_y=2$ 
for $\alpha=1$, and $M^{}_x=4$ and $M^{}_y=4$ for $\alpha=2$. In order to investigate the 
magnetic field dependence of the gap we fix the value of $\alpha$ and change the magnetic 
field $B$ and the period of the periodic potential simultaneously.

\begin{figure}
\includegraphics[width=8cm]{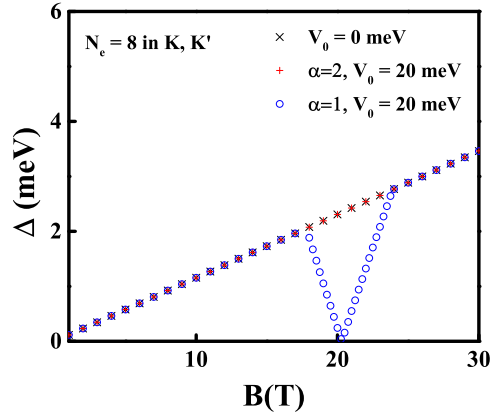}
\caption{\label{fig:Gap_Ne8TV} The gap between ground and first excited state of an 
eight-electron system in two valleys versus the magnetic field $B$ for two values of $\alpha$, 
with $V^{}_0=20$ meV and without the external periodic potential.}\end{figure}

In Fig.~\ref{fig:Gap_Ne8TV} the dependence of the gap between the ground state and the first excited 
state on the magnetic field strength is presented for $N^{}_e=8$ electrons in $K$ and $K'$ valleys 
for $\alpha=1$ and for $\alpha=2$ with $V^{}_0=20$ meV and without the periodic potential. In the 
absence of the periodic potential the change of $\alpha$ results only in the change of geometry 
of the system and therefore does not have any contribution in the gap value and in the figures the 
$V^{}_0=0$ case is presented without the indication of the value of $\alpha$. For $V^{}_0=0$ 
and for all values of the magnetic field the ground state corresponds to four electrons in each 
valley, all electrons having their spins in the opposite direction to the magnetic field. 
As for the excited state, it corresponds to the spin flip of one electron from eight 
electrons and the cases when the spin flip electron is located in the same valley as the 
partially filled spin down state with three electrons or they are located in two different 
valleys are degenerate. It should be also noted that there is no momentum transfer in this 
transition from the ground state to the excited state described above and the gap is equal to 
the Zeeman energy of the spin flip. Therefore for $V^{}_0=0$ the electron-electron 
interaction does not have any contribution in the lowest gap of the system. As can be seen 
from Fig.~\ref{fig:Gap_Ne8TV}, surprisingly the situation remains the same for $\alpha=2$ and 
$V^{}_0=20$ meV. 

\begin{figure}
\includegraphics[width=8cm]{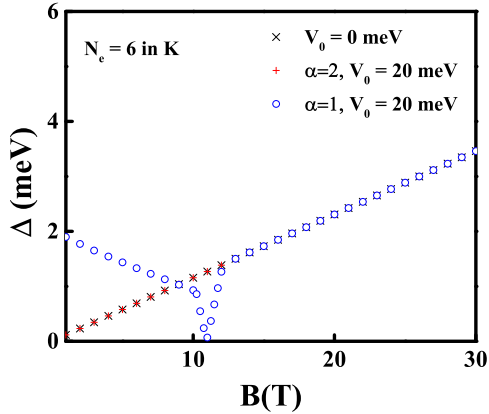}
\caption{\label{fig:Gap_Ne6} Same as in Fig.~\ref{fig:Gap_Ne8TV}, but for the system of six 
electrons. Only the $K$ valley is considered in this case.}\end{figure}

As for the case of $\alpha=1$ and $V^{}_0=20$ meV the situation is different. In the magnetic 
field region up to 18 Tesla the total spin of the ground state is $S^{}_z=-2$ and there are 
four electrons in each valley. This corresponds to the case when in each valley three electrons 
have spin down and one electron spin up. Therefore, the periodic potential changes the ground 
state of the system from the fully spin polarized state to the spin partially polarized state 
in this case. As for the excited state the total spin is $S^{}_z=-1$, which again means one 
additional spin flip and again the level is degenerate with respect to the exchange of spin 
up parts between the valleys. In Fig.~\ref{fig:Gap_Ne8TV} the gap is again equal to the Zeeman 
energy of the spin flip. At $B\approx18$ Tesla there is a crossing between the first and the
second excited states and up to $B=20$ Tesla the first excited state total spin is $S^{}_z=-4$, 
and the gap correspond to the transition between the states with double spin flip from the spin 
up to down and is again equal to the Zeeman energy of that transition. At $B\approx20$ Tesla 
there is a crossing between the ground and the first excited state, and after that the ground state 
is the state with total spin $S^{}_z=-4$. The first excited state is the state with total 
spin $S^{}_z=-2$ up to 24 Tesla and the state with total spin $S^{}_z=-3$ (which is again degenerate 
due to the different configurations of the spin up states between two valleys) afterwards. For 
these two parts, the gap is again equal only to the Zeeman energy for the appropriate transition. 
It should be noted that for the range of magnetic fields considered in this case both the 
ground and the excited state are described by the total momentum equal to zero and there 
is no momentum transfer in these transitions. Although the gap energy for this case is 
always equal to the Zeeman energy of the appropriate transition, the gap structure shown
for $\alpha=1$ and $V^{}_0=20$ meV is not the single-particle effect. Both the electron-electron 
interaction and the periodic potential are essential for the system to deviate from the
ferromagnetic state at low magnetic fields and afterwards for the observation of the 
transition from a spin partially polarized state to the fully polarized spin state by 
increasing the magnetic field.

\begin{figure}
\includegraphics[width=8cm]{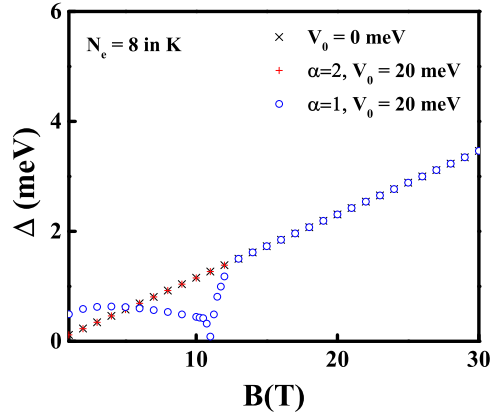}
\caption{\label{fig:Gap_Ne8} Same as in Fig.~\ref{fig:Gap_Ne6} but for the system of 
eight electrons. Only the $K$ valley is considered in this case.}\end{figure}

In Fig.~\ref{fig:Gap_Ne6} and Fig.~\ref{fig:Gap_Ne8} the dependence of the gap between 
the ground state and the first excited state on the magnetic field strength are presented 
for $N^{}_e=6$ and $N^{}_e=8$ electrons for $\alpha=1$ and $\alpha=2$ with $V^{}_0=20$ meV 
and without the periodic potential. Only the $K$ valley is considered in these cases. The 
cases of $V^{}_0=0$ and $V^{}_0=20$ meV with $\alpha=2$ show the same behavior as for the case 
of $N^{}_e=8$ electrons in $K$ and $K'$ valleys shown in Fig.~\ref{fig:Gap_Ne8TV}. The
ground state is the ferromagnetic state for all values of the magnetic field and the 
gap corresponds to the transition to the excited state with a single spin flip. For the 
case of $\alpha=1$ and $V^{}_0=20$ meV the situation is different. At small magnetic 
fields (up to 10 Tesla) the ground state total spin is equal to $S^{}_z=0$ (spin unpolarized state) 
and the transition corresponds to the excited state with total spin equal to $S^{}_z=-1$. 
It should be noted that in addition to the spin flip, there is also the momentum transfer in 
these transitions, because the first excited stated is characterized by a nonzero momentum 
up to 10 Tesla. Therefore these transitions correspond to collective excitations and the 
gap energy is comprised both the Zeeman term plus the $\Delta\propto\sqrt{B}$ term. This 
structure is clearly visible for the $N^{}_e=8$ electron case, but for the $N^{}_e=6$ electron 
case the collective nature of the excitation is completely suppressed by the Zeeman term 
for magnetic fields considered in Fig.~\ref{fig:Gap_Ne6}. In Fig.~\ref{fig:Gap_Ne6} 
and Fig.~\ref{fig:Gap_Ne8}, starting with $B=10$ Tesla the structure of the transition is the same 
as for the case of $N^{}_e=8$ electrons in $K$ and $K'$ valleys shown in Fig.~\ref{fig:Gap_Ne8TV}. 
There are several crossings between the low-lying excited states and the ground states 
of the system passes from the spin unpolarized ground state to the ferromagnetic (fully 
polarized) state. In the ferromagnetic regime the gap again corresponds to the spin flip 
and is equal to the Zeeman energy of that flip.    

We now use the features observed in this work to interpret the result shown in Fig. 4 
of Ref.~\cite{geim_14} for the filling factor $\nu=0$. For magnetic fields up to 20 Tesla, due to 
the valley anisotropic terms \cite{Abanin} and also due to the spin unpolarized state 
observed for low magnetic fields in our work, presumably the system is in a spin unpolarized
state. The almost linear dependence of the gap at magnetic fields up to around 5 Tesla 
and the $\sqrt{B}$ dependence for magnetic fields between 5-20 Tesla indicates that the 
excitations have both spin flip and the momentum transfer component (collective excitation). 
Based on these observations it can be  assumed that there is a competition between these two 
components and that for up to 5 Tesla magnetic fields the Zeeman contribution is dominant in 
the gap energy, whereas in the range of magnetic fields 5-20 Tesla the electron-electron 
interaction contribution is dominant. The lowering of the gap in the region close to $\alpha=1$, 
then almost linear dependence after around 28 Tesla and the absence of similar features for 
$\alpha=2$ closely resembles the results obtained in our paper. We therefore suggest that this 
behavior indicates that there is a phase transition around $\alpha=1$, i.e., there is a transition 
from the spin unpolarized state to the partially or fully spin polarized state.

In conclusion, we have considered the influence of the periodic potential due to the 
moir\'e lattice on the dependence of the energy gap on the magnetic field for filling 
factor $\nu=0$ in graphene using the exact diagonalization scheme. We have considered three 
cases: $N^{}_e=8$ electrons in $K$ and $K'$ valley, $N^{}_e=6$ and $N^{}_e=8$ electrons located 
only in the $K$ valley and disregarding the contribution of the $K'$ valley. In all cases we 
find that for $\alpha=1$, inclusion of the periodic potential and the electron-electron interactions 
results in unpolarized or partially polarized ground state at lower magnetic fields and a transition 
to the fully polarized ground state by increasing the magnetic field. This behavior is 
not observed for the $\alpha=2$ case, where we only observe the ferromagnetic state for 
all values of the magnetic field. The similarity between the results obtained here and reported
in Fig. 4 of Ref.~\cite{geim_14} was analyzed and a possible explanation was presented 
on the behavior of the dependence of the gap on the perpendicular magnetic field observed 
in that work.          

The work has been supported by the Canada Research Chairs Program of the 
Government of Canada.


\begin{thebibliography}{99}
\bibitem[\ddag]{byline} Electronic address:
Tapash.Chakraborty@umanitoba.ca

\bibitem{graphene_book}
H. Aoki and M.S. Dresselhaus (Eds.), {\it Physics of Graphene}
(Springer, New York 2014).

\bibitem{abergeletal}
D.S.L. Abergel, V. Apalkov, J. Berashevich, K. Ziegler, and
T. Chakraborty, Adv. Phys. {\bf 59}, 261 (2010).

\bibitem{FQHE_chapter}
T. Chakraborty and V. Apalkov, in \protect\cite{graphene_book} Ch. 8;
T. Chakraborty and V.M. Apalkov, Solid State Commun. {\bf 175}, 123 (2013).

\bibitem{interaction}
V. Apalkov and T. Chakraborty, Solid State Commun. {\bf 177}, 128 (2014);
D.S.L. Abergel and T. Chakraborty, Phys. Rev. Lett. {\bf 102}, 056807 (2009);
D. Abergel, V. Apalkov, and T. Chakraborty, Phys. Rev. B {\bf 78}, 193405 (2008);
D. Abergel, P. Pietil\"ainen, and T. Chakraborty, Phys. Rev.B {\bf 80}, 081408
(2009); V. Apalkov and T. Chakraborty, Phys. Rev. B {\bf 86}, 035401 (2012).

\bibitem{mono_FQHE} V.M. Apalkov and T. Chakraborty, Phys. Rev. Lett.
{\bf 97}, 126801 (2006).

\bibitem{bi_FQHE}
V.M. Apalkov and T. Chakraborty, Phys. Rev. Lett. {\bf 105}, 036801 (2010);
Phys. Rev. Lett. {\bf 107}, 186803 (2011).

\bibitem{FQHE_expt}
X. Du, I. Skachko, F. Duerr, A. Luican, and E.Y. Andrei, Nature {\bf 462}. 192
(2009); D.A. Abanin, I. Skachko, X. Du, E.Y. Andrei, and L.S. Levitov, Phys.
Rev. B {\bf 81}, 115410 (2010); K.I. Bolotin, F. Ghahari, M.D. Shulman,
H.L. St\"ormer, and P. Kim, Nature {\bf 462}, 196 (2009); F. Ghahari, Y. Zhao,
P. Cadden-Zimansky, K. Bolotin, and P. Kim, Phys. Rev. Lett. {\bf 106},
046801 (2011).

\bibitem{langbein}
D. Langbein, Phys. Rev. {\bf 180}, 633 (1969). 

\bibitem{hofstadter}
D. Hofstadter, Phys. Rev. B {\bf 14}, 2239 (1976).

\bibitem{graphene_butterfly}
T. Chakraborty and V.M. Apalkov, arXiv:1408.4485 (2014).

\bibitem{dean_13}
C.R. Dean, L. Wang, P. Maher, C. Forsythe, F. Ghahari, Y. Gao, J. Katoch,
M. Ishigami, P. Moon, M. Koshino, T. Taniguchi, K.Watanabe, K.L. Shepard,
J.Hone, and P. Kim, Nature {\bf 497}, 598 (2013).

\bibitem{hunt_13}
B. Hunt, J.D. Sanchez-Yamagishi, A.F. Young, M. Yankowitz, B.J. LeRoy,
K. Watanabe, T. Taniguchi, P. Moon, M. Koshino, P. Jarillo-Herrero,
and R.C. Ashoori, Science {\bf 340}, 1427 (2013).

\bibitem{geim_13}
L.A. Ponomarenko, R.V. Gorbachev, G.L. Yu, D.C. Elias, R. Jalil, A.A. Patel,
A. Mishchenko, A.S. Mayorov, C.R. Woods, J.R. Wallbank, M. Mucha-Kruczynski,
B.A. Piot, M. Potemski, I.V. Grigorieva, K.S. Novoselov, F. Guinea, V.I. Falko
and A.K. Geim, Nature {\bf 497}, 594 (2013).

\bibitem{apalkov_14}
V.M. Apalkov and T. Chakraborty, Phys. Rev. Lett. {\bf 112}, 176401 (2014).

\bibitem{geim_14}
G.L. Yu, R.V. Gorbachev, J.S. Tu, A.V. Kretinin, Y. Cao, R. Jalil, F. Withers,
L.A. Ponomarenko, B.A. Piot, M. Potemski, D.C. Elias, X. Chen, K. Watanabe,
T. Taniguchi, I.V. Grigorieva, K.S. Novoselov, V.I. Falko, A.K. Geim, and
A. Mishchenko, Nat. Phys. {\bf 10}, 525 (2014).

\bibitem{areg_butterfly}
A. Ghazaryan, T. Chakraborty, and P. Pietil\"ainen, arXiv:1408.3424 (2014).

\bibitem{mag_translation}
J. Zak, Phys. Rev. {\bf 133}, A1602 (1964); E. Brown, Phys. Rev. {\bf 133},
A1038 (1964)

\bibitem{haldane_85}
F.D.M. Haldane, Phys. Rev. Lett. {\bf 55}, 2095 (1985).

\bibitem{read}
A. Kol and N. Read, Phys. Rev. B {\bf 48}, 8890 (1993).

\bibitem{FQHE_book}
T. Chakraborty and P. Pietil\"ainen, {\it The Quantum Hall Effects} (Springer, New York
1995); T. Chakraborty, and P. Pietil\"ainen, {\it The Fractional Quantum Hall Effect}
(Springer, New York 1988).

\bibitem{review}
U. R\"ossler and M. Shurke, in {\it Advances in Solid State Physics},
edited by B. Kramer (Springer, Berlin 2000), Vol. 40, pp. 35-50.

\bibitem{vidar}
V. Gudmundsson and R.R. Gerhardts, Surf. Sci. {\bf 361-362}, 505 (1996);
Phys. Rev. B {\bf 52}, 16744 (1995); Phys. Rev. B {\bf 54}, 5223R (1996).

\bibitem{ando}
M. Koshino and T. Ando, J. Phys. Soc. Jpn {\bf 73}, 3243 (2004).

\bibitem{note}
The periodic rectangular geomery was extensively used earlier in the study of the
FQHE in various situations. For example, see
T. Chakraborty, Surf. Sci. {\bf 229}, 16 (1990); Adv. Phys. {\bf 49}, 959 (2000);
T. Chakraborty and P. Pietil\"ainen, Phys. Rev. Lett. {\bf 76}, 4018 (1996);
T. Chakraborty and P. Pietil\"ainen, Phys. Rev. Lett. {\bf 83}, 5559 (1999);
T. Chakraborty and P. Pietil\"ainen, Phys. Rev. B {\bf 39}, 7971 (1989);
V.M. Apalkov, T. Chakraborty, P. Pietil\"ainen, and K. Niemel\"a, Phys. Rev. Lett.
{\bf 86}, 1311 (2001); T. Chakraborty, P. Pietil\"ainen, and F.C. Zhang, Phys. Rev.
Lett. {\bf 57}, 130 (1986); T. Chakraborty and F.C. Zhang, Phys. Rev. B {\bf 29},
7032 (R) (1984); F.C. Zhang and T. Chakraborty, Phys. Rev. B {\bf 30}, 7320 (R) (1984).

\bibitem{Abanin}
D.A. Abanin, B.E. Feldman, A. Yacoby and B.I. Halperin, Phys. Rev. {\bf 88}, 115407 (2013). 

\end{thebibliography}
\end{document}